# Kondo Impurities in Two Dimensional MoS$_2$ for Achieving Ultrahigh Thermoelectric Powerfactor

*Jing Wu* [1,4†], *Yanpeng Liu* [2,4†], *Yi Liu* [3,4], *Yongqing Cai* [5], *Yunshan Zhao* [3,4], *Hong Kuan Ng* [1,6], *Kenji Watanabe* [7], *Takashi Taniguchi* [7], *Gang Zhang* [5], *Chengwei Qiu* [3,4,8], *Dongzhi Chi* [1], *AH Castro Neto* [4,5], *John TL Thong* [3,4\*], *Kian Ping Loh* [2,4\*], *Kedar Hippalgaonkar* [1,4\*]

[1] Institute of Materials research and Engineering, 2 Fusionopolis Way, Innovis, #08-03, Agency for Science, Technology and Research, 138634, Singapore

[2] Department of Chemistry, National University of Singapore, 117542, Singapore

[3] Department of Electrical and Computer Engineering, National University of Singapore, 117583, Singapore

[4] Centre for Advanced 2D Materials and Graphene Research Centre, National University of Singapore, 117546, Singapore

[5] Institute of High Performance Computing, Agency for Science, Technology and Research, 138632, Singapore

[6] Department of Physics, National University of Singapore, 117542, Singapore

[7] National Institute for Materials Science, Tsukuba, Ibaraki, 305-0044, Japan

[8] SZU-NUS Collaborative Innovation Center for Optoelectronic Science and Technology, Shenzhen University, Shenzhen, 518060, China

†These authors contributed equally to this work.

\* Email: kedarh@imre.a-star.edu.sg (K.H.); chmlohkp@nus.edu.sg (K.P. L.); elettl@nus.edu.sg (J.T.L.T.)

**Local magnetic impurities arising from atomic vacancies in two-dimensional (2D) nanosheets are predicted to have a profound effect on charge transport due to resonant scattering, and provide a handle for enhancing thermoelectric properties through the Kondo effect. However, the effects of these impurities are often masked by external fluctuations and turbostratic interfaces, therefore, it is highly challenging to probe the correlation between magnetic impurities and thermoelectric parameters experimentally. In this work, we demonstrate that by placing Molybdenum Disulfide ($MoS_2$) on a hexagonal Boron Nitride ($h$-BN) substrate, a colossal spin splitting of the conduction sub-band up to ~50.0±5.0 meV is observed at the sulfur vacancies, suggesting that these are local magnetic states. Transport measurements reveal a large anomalous positive Seebeck coefficient in highly conducting $n$-type $MoS_2$, originating from quasiparticle resonance near the Fermi level described by the Kondo effect. Furthermore, by tuning the chemical potential, a record power factor of 50mW/m·K$^2$ in low-dimensional materials was achieved. Our work shows that defect engineering of 2D materials affords a strategy for controlling Kondo impurities and tuning thermoelectric transport.**

Thermoelectrics are solid state energy converters that can be used to harvest electrical energy from waste heat, thus they are attractive as sustainable energy resource. The performance of thermoelectric materials is characterized by the thermoelectric figure of merit (ZT, defined as $S^2\sigma T/\kappa$): the key bottleneck is the anti-correlation between the Seebeck coefficient ($S$) and electrical ($\sigma$)/thermal ($\kappa$) conductivities. An ideal thermoelectric material should exhibit simultaneously large $S$ and $\sigma$.[1] It has been proposed that two-dimensional (2D) materials, with their discretized density of states (strong quantum confinement), may show better performance in thermoelectrics compared to bulk materials.[2] For instance, SnSe has been found to have a figure of merit (ZT) of 2.8±0.5. The remarkably high ZT is due to the anharmonicity of its chemical bonds, giving it an ultralow thermal conductivity at high temperature (700-900K).[3,4] However the power factor (PF = $S^2\sigma$) is still relatively modest (1 mW/m·K$^2$).[3] New transport mechanisms are needed to push the PF of thermoelectric materials beyond the well-known $S$-$\sigma$ anti-correlation limit.[1] For example, the violation of the Mott relation results in enhanced PF in the hydrodynamic charge transport regime in graphene due to strong electron-electron correlations.[5] It is well known that magnetic impurities not only strongly couple with itinerant charge carriers but also significantly affect and even reverse charge transport behavior.[6–8] The atomic thickness of 2D materials renders it highly sensitive to such extrinsic effects, and therefore they serve as a good platform for exploiting thermoelectric properties by introducing magnetic impurities. Among these materials, MoS$_2$ has attracted special attention because of its well-defined spin-splitting under light illumination and/or applied magnetic fields.[9–11] Extrinsic magnetic impurities in MoS$_2$ are predicted to be introduced by vacancies, dislocations, edges, strain or doping by magnetic ions.[12–15] However, due to its extreme sensitivity to external fluctuations and turbostratic interfaces, there has not yet been a clear demonstration of how the thermoelectric properties of 2D materials can be influenced by magnetic impurities.

In this work, we report that when few-layer MoS$_2$ flakes are interfaced to hexagonal boron nitride ($h$-BN) substrate, the correlation between charge carriers and magnetic impurities in MoS$_2$ is clearly observed due to reduced scattering by charged surface states, impurities and surface roughness. Using low-temperature scanning tunneling microscopy (LT-STM), single-atom sulfur vacancies (up to ~1.81±0.4×10$^{12}$ cm$^{-2}$) were observed on MoS$_2$. The conduction sub-band near the vicinity of any such sulfur vacancy shows a colossal band splitting of ~50±5 meV. Density Functional Theory (DFT) simulations reveal that this splitting originates from sulfur-vacancy-induced magnetic states. When fabricated into a MoS$_2$/$h$-BN field effect transistor (FET) device, we observed an anomalous sign change of the Seebeck coefficient and an extremely large positive Seebeck value (2 mV/K) even in the metallic regime. In accordance with the revised Boltzmann transport equation (BTE), the exceptional thermoelectric transport behavior of MoS$_2$/$h$-BN device is attributed to the strong interaction of the electrons with these local magnetic impurities *via* Kondo resonance at the Fermi level. In addition, tuned by an external electrostatic field, the PF of MoS$_2$/*h*-BN device approaches a new record of 50 mW/m·K$^2$. Our results demonstrate that vacancy-rich MoS$_2$ crystals are a great material platform for next generation thermoelectric and energy applications.

**Vacancy induced magnetic moment**s

Inversion symmetry breaking at vacancy or edge sites of MoS$_2$ has been theoretically predicted to enable local magnetic moments and induces spin splitting of both valence and conduction bands.[16] To probe spin splitting induced by vacancies in MoS$_2$, an atomically flat and high dielectric *h*-BN as substrate is needed to avoid spurious effect from impurities at the interfaces. The MoS$_2$/h-BN heterostructure device was fabricated by a well-established transfer technique and then loaded into the LT-STM chamber and annealed in ultrahigh vacuum ($< 1 \times 10^{-10}$ torr) to ensure better electrical and thermal contact and a cleaner surface. (more details in Supplementary Note S1 and Note S2 ).

**Fig. 1**a shows a typical STM topographic image of MoS$_2$ flake on *h*-BN for a scan area of 75 nm × 75 nm. The thickness of MoS$_2$ flakes was determined to be six layers using atomic force microscopy. Pristine MoS$_2$ lattice is observed in the majority of the scanned regions. The dark topographic contrast in the STM image (Fig. 1a) and lattice discontinuity are hallmarks of single sulfur vacancies,[17] with a concentration of ~1.81±0.4×10$^{12}$ cm$^{-2}$. In the high-resolution image inserted in Fig. 1c (center panel), a hexagonal lattice of 3.15±0.5 Å periodicity is seen (Supplementary Fig. S3), corresponding to the atomic lattice of the top sulfur layer in MoS$_2$ crystal.[18] The defects are experimentally observed at the center of sulfur lattice sites, exclude other defect types such as interstitials, Mo vacancy or antisite defects (a Mo atom substituting a S$_2$ column or vice versa).[19] Scanning tunneling spectroscopy (STS), which probes the local density states, was applied to study the effect of such sulfur vacancy defects on the electronic properties. To better resolve the effect of sulfur vacancies, a lower set point ($V_S$ = -0.7V, *I* = 1.3nA) was used to collect the *d*I/*d*V

spectrum from the vicinity of sulfur vacancy as well as far away from it (Fig. 1c). At the pristine MoS$_2$ region (center panel, Fig. 1c), a series of oscillation peak/dip features were observed at the conduction band region and their spacings at ~98.0 meV, ~134.3 meV and ~169.0 meV are in good agreement with the predicted sub-band ($C_1$, $C_2$ and $C_3$, right panel of Fig. 1c) structure of six-layer MoS$_2$. When the STM tip is located on top of a vacancy, a splitting (energy ~50.0±5.0 meV) of these sub-bands was clearly captured.

A survey X-ray photoelectron spectroscopy scan over the whole MoS$_2$ flake shows no evidence of magnetic elemental impurities (Supplementary Fig. S4). In order to investigate the origin of sub-band splitting in vacancy-rich MoS$_2$ crystals, first-principle calculations were conducted (more calculation details in Supplementary Note S4). The removal of one sulfur atom in the MoS$_2$ sheet creates a single sulfur vacancy ($V_{Sul}$), accompanied by dangling states and exposed Mo atoms in the vacancy core. Single neutral $V_{Sul}$ has two states: a fully occupied singlet $A$ state and an empty doubly degenerate $E$ state.[20] When carriers are injected into the MoS$_2$, the $V_{Sul}$ state becomes negatively charged and the $V_{Sul}^{-1}$ is spin-polarized, accompanied by a spin moment of $\mu=1/2$. The spin density associated with $V_{Sul}^{-1}$, with a significant component localized at the three exposed Mo atoms, is shown in Fig. 1 d&e. Charging *via* gating alters the local magnetic moment and state-splitting of the $V_{Sul}$, therefore, allowing tuning of the scattering of charge carriers, especially at low temperatures.[21]

**Electrical performance due to magnetic vacancies**

To investigate the influence of vacancy-induced magnetic impurities, low temperature transport measurements were carried out. **Fig. 2**a shows the top- and section-view of the MoS$_2$/*h*-BN heterostructure and a representative optical image is shown in Fig. 2b. To be consistent, MoS$_2$ flake with thickness ~4.5 nm was selected (six layers, Fig. 1c). Four-probe measurements were performed to exclude the effects of electrical contact resistance and the nanofabricated heater allows thermoelectric measurements. The linear $I_{sd}$-$V_{sd}$ curves collected at room temperature (Fig. 2c) indicate ohmic contacts between metal electrode and $MoS_2$ flake (Supplementary Fig. S5). The carrier concentration ($n$) in $MoS_2$ is modulated by applying a back-gate voltage ($V_g$). Fig. 2d. shows conductance increasing with $V_g$ for the $MoS_2$/$h$-BN sample for measured temperature ($T$) range, typical $n$-type FET behavior. From 300K to 100K, a clear crossing point (at $V_g \sim 20V$) appeared, indicating that $MoS_2$ undergoes a routine metal-to-insulator transition (MIT).[22] At $T$ <100K, the conductance drops anomalously as $T$ decreases, in contrast with the trend exhibited by $MoS_2$ on Si/$SiO_2$ devices. This is explained by the strong coupling between magnetic impurities and conduction electrons, and will be discussed later.

To better visualize the correlation between magnetic impurities and conduction electrons in $MoS_2$/$h$-BN, temperature-dependent four-probe conductance ($G$) were plotted as a function of $V_g$ for both $MoS_2$/$h$-BN (**Fig. 3**a) and $MoS_2$/$SiO_2$ (Fig. 3b) devices for comparison. In the high temperature range ($T$ >100K), and close to $G_{MIT} \sim e^2/h$, the second order electron correlation driven MIT can be observed clearly in both devices. For $MoS_2$/$SiO_2$ device, such behavior exists over the temperature range from 300K to 20K. However, for the $MoS_2$/$h$-BN device, instead of saturating at a residual value as $T \rightarrow 0$, a conductance peak (temperature at this critical point is defined as $T_{max}$, red dashed line in Fig. 3a) in the metallic region is observed. This observed anomalous resistance minimum at low temperature is a signature of the Kondo effect, which is induced by the magnetized sulfur vacancies.

The Kondo effect is suppressed in the $MoS_2$/$SiO_2$ device due to the imperfect interface between $MoS_2$ and $SiO_2$, but clearly detectable for the $MoS_2$/$h$-BN device. For a Kondo system, the magnetic impurity undergoes a spin-flip process and electrons at the Fermi level can quantum-mechanically tunnel from the magnetic impurity. This spin-flip Kondo scattering, dominates charge transport in this temperature range and gives rise to the anomalous component of the resistance. The signature of Kondo-driven transport could also be seen in the change in four-probe mobility ($\mu$) with temperature (Fig. 3c, calculation detail in Supplementary Note S6). For $MoS_2$-on-$SiO_2$, $\mu$ is mainly limited by optical phonon scattering and its temperature dependence follows a power law of $\mu \sim T^{-\gamma}$ when $T$> 100K.[23] Once $T$<

100K, most optical phonons are de-activated and saturation is observed, resulting in $d\mu/dT \sim 0$ as $T\rightarrow 0$.[23] In contrast, MoS$_2$/h-BN device deviates from this trend, the mobility starts to decrease as $T <100$ K and shows $d\mu/dT > 0$ behavior, which confirms the existence of the Kondo scattering at low temperature.

Fig. 3d shows the four-probe resistance plots of MoS$_2$/h-BN device at three different back-gates. For each curve, a clear resistance minimum can be observed at 70K, 89K and 135K, respectively. For a better understanding of these data, apart from 2D Block-Gruneisen resistance ($\propto T^4$) [24,25] and electron-phonon high temperature resistance ($\propto T$), the Kondo resistance ($R_K$, full expression in Supplementary Note S11) also need to be taken into account:

$$R = AT^4 + BT + R_K + R_0 \tag{1}$$

where $R_0$ is the temperature independent term arising from a residual zero temperature resistance. The well-fitted curves in Fig. 3d confirmed the contribution from Kondo resistance. At certain temperature region (Kondo temperature, $T_K$), magnetic impurities quantum-mechanically exchange spin with conduction electrons of many wave-vectors (momenta)[26], creating a resonant scattering state at the Fermi Level with a width $\sim k_B T_K$ [$k_B T_K \sim \Delta \exp(\frac{-1}{D(E_F)J_0})$, where $D(E_F)$ is the electron density of states (DOS) at the Fermi level, $\Delta$ is the bandwidth and $J_0$ is a constant representing a Kondo scattering exchange energy].[27,28] As a result, a dip in the resistance vs temperature curve appear. To further exploit the DOS dependence of Kondo temperature, Fig. 3e summaries the $T_K$ variation as function of $V_g$ for MoS$_2$/h-BN sample. At $V_g > V_{MIT}$, $T_K$ remains nearly unchanged because of the constant 2D density of states. When $V_g < V_{MIT}$, the Fermi level shifts toward the gap region, therefore, $T_K$ starts to increase with lower doping; similar to the behavior observed in heavily- doped fermion bulk systems such as CePd$_3$.[29] We also scale the $R(T)$ curves at each $V_g$ and observe a universal Kondo behavior,[6,30] in which the normalized Kondo resistance $R_k/R_{k_0}$ vs $T/T_k$ at all gate voltages collapses on to a single universal Kondo behavior curve (Fig. 3h). This is in accordance with Numerical

Renormalization Group theory,[31] which attests to the Kondo-driven transport mechanism of $MoS_2$/$h$-BN devices.

**Thermoelectric performance driven by Kondo effect**

To investigate the influence of magnetic impurities on the thermoelectric performance, a DC current is introduced *via* a nano-fabricated heater to create a temperature gradient along the devices (Fig. 2a). **Fig. 4**a shows the Seebeck coefficient (*S*) as a function of temperature at $V_g$= 70V, 50V and 30V for both $MoS_2$/$h$-BN and $SiO_2$ devices. The absolute |*S*| decreases with increasing carrier concentration in the $MoS_2$ channel (details in Supplementary Note S13 and Fig. S15), comparable to the reported trends in earlier work.[32,33] Both $MoS_2$/$SiO_2$ and $MoS_2$/$h$-BN sample show negative *S* behavior for *T* > 120K. However, at even lower temperatures, an anomalous sign-change (from negative to positive) for *S* is observed for $MoS_2$/$h$-BN even though electrons remain as the dominant charge carriers. The positive *S* reaches as high as 2mV/K for $V_g$=30V, a remarkably large value for a material in the metallic state. This anomalous change in the sign of *S* proves that magnetic impurities embedded in $MoS_2$ strongly scatter the diffusive electrons due to the temperature gradient and change the thermoelectric transport completely. To better illustrate the sign-change of *S*, we plot a 2D map of *S* as a function of both *T* and $V_g$ in Fig. 4b. Clearly, *S* changes from negative to positive values at all gate voltages. It is also found that the critical temperature (defined as the temperature at which *S* =0) and the maximum positive *S* (highest intensity, red in Fig. 4b) increases as $V_g$ decreases, suggesting that the Seebeck coefficient can be tuned by electrostatic gating. Importantly, this line for *S*=0 is identical to the conduction maximum line $T_{max}$ in Fig. 3a, further evidencing that this anomalous thermoelectric transport originates from Kondo scattering.

Since the transfer curve shows a clear *n*-type behavior and the quasiparticle bandgap of 6-layer $MoS_2$ is relatively large,[34] the valence band states of $MoS_2$ are inaccessible, especially at low temperatures with positive *S*. When electrons are the sole carriers, the single-particle Boltzmann transport equation (BTE) could be employed to describe thermoelectric transport. In the degenerate state (highly conducting on-state), the BTE reduces to the Mott formula described below[35]:

$$S = -\frac{\pi^2}{3}\frac{k_B^2 T}{|e|}\left[\frac{\partial \ln\tau}{\partial E} + \left(\frac{\partial \ln g}{\partial E} + \frac{\partial \ln v^2}{\partial E}\right)\right]\bigg|_{E=E_F} = S_\tau + S_N \qquad (2)$$

with $S_\tau = -\frac{\pi^2}{3}\frac{k_B^2 T}{|e|}\frac{\partial ln\tau}{\partial E}\bigg|_{E=E_F}$ and $S_N = -\frac{\pi^2}{3}\frac{k_B^2 T}{|e|}\left(\frac{\partial \ln g}{\partial E} + \frac{\partial \ln v^2}{\partial E}\right)\bigg|_{E=E_F}$.

Here, $\tau$, $g$ and $v$ indicates the relaxation time, density of states and group velocity, respectively. $S_\tau$ and $S_N$ are the contribution arising from the energy dependent scattering and diffusion. The sign of $S_N$ only depends on the charge carrier type, thus *n*-type materials exhibit a negative $S_N$. For 2D system dominated by acoustic phonon scattering, $\tau$ is energy independent.[36] Hence, the *S* of our $MoS_2/SiO_2$ is dominated by negative $S_N$. In accordance to Equation (2), an overall positive *S* results when $S_\tau > 0$ (only when $\frac{\partial \ln\tau}{\partial E} < 0$) for *n*-type materials. From the Seebeck coefficient and mobility at $V_g$=70V (Fig. 4c), it is found that the mobility gives a peak value right at zero *S*.

In the Kondo physics framework, scattering is highly energy-dependent and the mobility is known to be proportional to $1/\tau$,[37] which is opposite to the conventional one-band system where $\mu \propto \tau$. Approximating $1/\partial E$ as $1/(k_B \partial T)$ [38] in $S_\tau$, $\frac{\partial \ln\tau}{\partial T} \propto -\frac{d\mu}{dT}$ could be derived to describe the relationship between Seebeck coefficient and mobility. From Fig. 4c, we can see that the measured *S* is positive at the region with $\frac{d\mu}{dT} < 0$, in good agreement with Kondo physics model. Fig. 4d illustrates the Kondo-driven thermoelectric transport more vividly. In a conventional 2D system ($MoS_2/SiO_2$, here), the electrons diffuse from the hot side to the cold side driven by the temperature gradient. In the case of $MoS_2/h$-BN device however, the interaction between electrons at Fermi energy and magnetic impurities strongly influences the non-equilibrium energy spectrum and as-generated Kondo resonance (a spin excitation state located at the Fermi level), this causes more conduction electrons to be scattered back and accumulate at the hot side, thus reversing the sign of Seebeck coefficient.

To determine the Kondo-scattered Seebeck coefficient ($S_\tau$), the two band approach that originates from the Hirst model is adopted.[39] $S_N$ is calculated by solving the linearized Boltzmann transport equation (BTE) under a relaxation time approximation(detailed calculation of $S_\tau$ and $S_N$ are provided in Supplementary Note S12).[32,40] Consequently, the overall Seebeck coefficient ($S_{total_\tau} = S_N + S_\tau$) can be determined. **Fig. 5**a shows the measured Seebeck coefficient as a function of temperature of MoS$_2$/$h$-BN device at carrier concentration of $n= 2\times10^{12}$ cm$^{-2}$, which is well captured by this Kondo scattering model (more cases in Supplementary Fig. S14). For higher gate voltages, the Fermi level of MoS$_2$/$h$-BN sample is pushed further from the impurity state, this causes Kondo scattering to be present only at a lower temperature (Fig. 4b). Due to the strong interaction of electrons with magnetic impurities through the Kondo effect, the larger $S$ values in the on-state of MoS$_2$ leads to an unusually strong enhancement in the thermoelectric PF ($S^2\sigma$). Fig. 5b shows the PF as a function of temperature for the on-state gate voltages. A high PF value of 50 mW/m·K$^2$, originating from Kondo-enhanced $S$, can be achieved, thus setting a new record for the thermoelectric power factor(Fig. 5c)[32,33,41–46].

In summary, we discover that magnetic impurities due to sulfur vacancies exert strong influence on the thermoelectric properties of few-layer MoS$_2$ when $h$-BN is used as the substrate, whereas these effects are suppressed on substrate such as SiO$_2$. The electric and thermoelectric transport in few-layer MoS$_2$ on $h$-BN exhibits a clear magnetically induced Kondo behavior, leading to a large anomalous positive Seebeck coefficient of 2 mV/K in $n$-type MoS$_2$ in the on-state. Importantly, our work shows that magnetic impurities-induced Kondo effect can be electrostatically tuned to manipulate the Seebeck coefficient and PF. The ability to exhibit both negative Seebeck (diffusive) and positive Seebeck (Kondo) coefficients in $n$-type MoS$_2$ suggest that a single doped material can be used to fabricate thermoelectric circuit devices, thus bypassing the need for complex p-n heterostructure device circuit. This work also points to the possibility of dopant engineering to make Kondo lattices[47] for nano-circuits in thermoelectric devices.

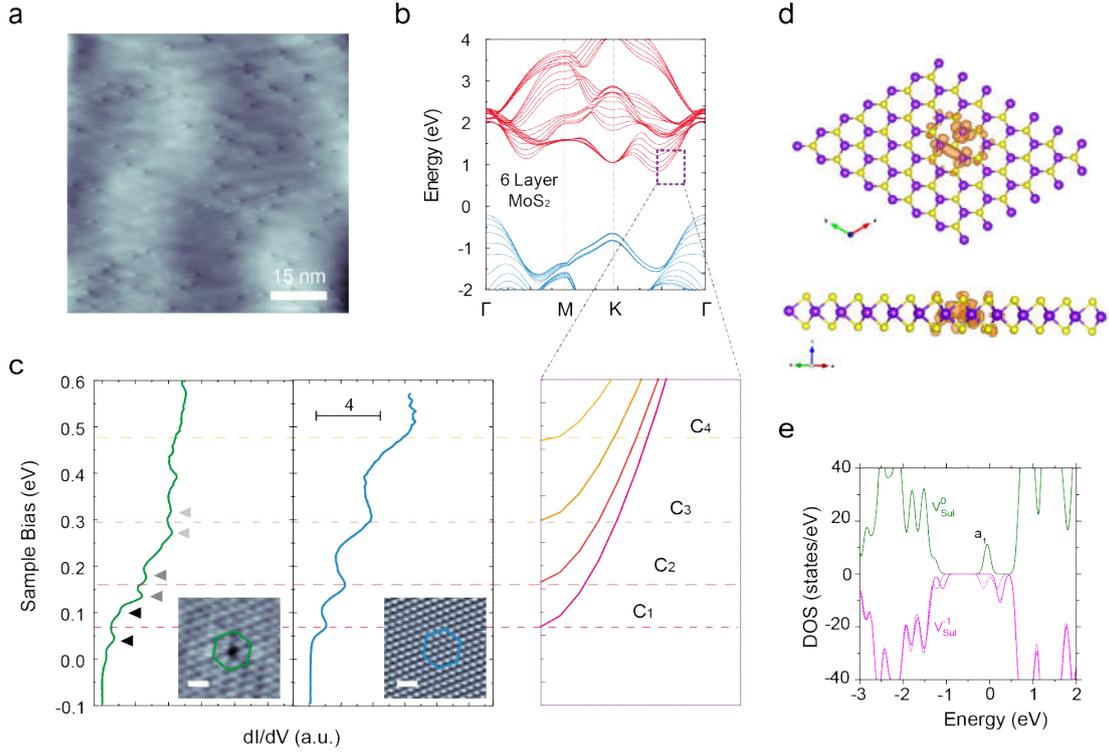

**Figure 1 | Electronic and magnetic properties of 6-layers MoS$_2$ with sulfur-vacancy. a**, Large-area STM image of the bare MoS$_2$ on *h*-BN substrate ($V_S$ = 1.0V, $I$ = 0.7nA). The dark topographic contrast, which shows the single sulfur vacancy features can be found over the whole scanned area. **b**, DFT calculation of 6-layer MoS$_2$. The zoomed-in image shows the conduction sub-bands ($C_n$) near the CBM at Λ point of 6-layer MoS$_2$. **c**, $dI/dV$ spectrum ($V_S$ = -0.7V, $I$ = 1.3nA) at the vicinity of a sulfur vacancy (scale bar 0.6 nm) and pristine region (scale bar 0.8nm) of MoS$_2$ near CBM. Conduction sub-bands $C_1$, $C_2$ and $C_3$ can be revealed from the resonance peaks in STS curve of pristine MoS$_2$. **d**, Top and side views of spin density associated with single sulfur vacancy. (purple larger spheres represent Molybdenum atoms and smaller yellow spheres represent Sulfur atoms). The spin density associated with $V_{Sul}^{-1}$, with a significant component localized at the three exposed Mo atoms. **e,** Electronic DOS of the sulfur vacancy in neutral ($V_{Sul}^{0}$) and -1 charged state ($V_{Sul}^{-1}$) (For intuitive, the VBMs for two cases was aligned).The charging of the vacancy leads to the lifting of the degeneracy of the a$_1$ defective level indicated by the splitting of curves of spin up (dashed line) and spin down (solid line) states for $V_{Sul}^{-1}$.

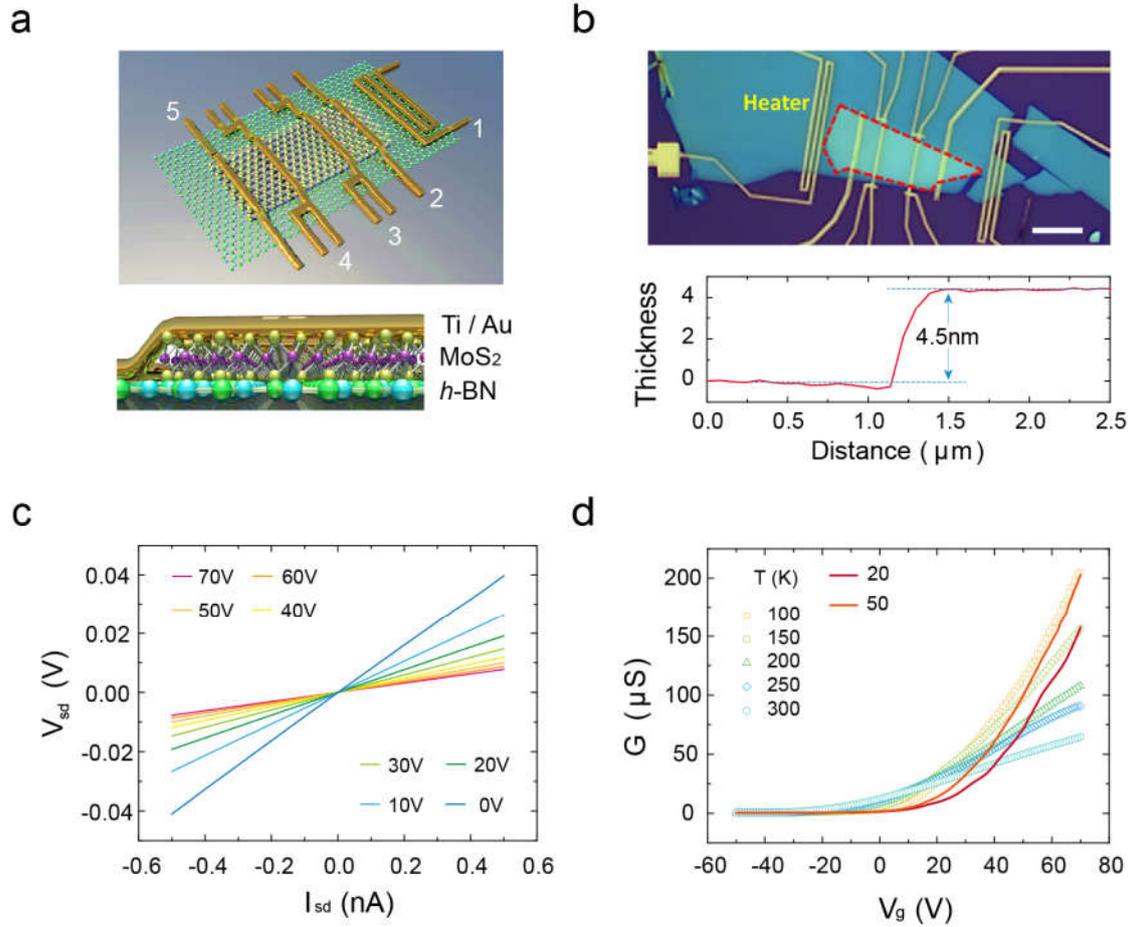

**Figure 2 | Structural and electronic properties of MoS$_2$/h-BN heterostructure. a**, Schematic diagram of the device. Electrode 1 acts as an electrical heater. Electrodes 2 and 5 act as a current source for 4-probe electrical measurements, while electrodes 3 and 4 act as thermometers. Bottom is the section-view of the heterostructure. **b**, Optical image of a complete device. The dotted red dashed box outlines the MoS$_2$ flake. The scale bar is 10 μm. **c**, 4-probe $I_{sd}$-$V_{sd}$ curves at different $V_g$ values at 300 K. **d**, 4-probe field effect transistor characterization at different temperatures. Clear crossover can be found at $V_g$~20V for $T$ >100K, which is defined as the Metal-to-Insulator Transition (MIT). For $V_g$> $V_{MIT}$, MoS$_2$ shows metallic behavior, conductance decreases with increasing $T$. For $V_g < V_{MIT}$, conductance increases with increasing $T$, which is a typical insulating behavior. When $T <$ 100K, the conductance (in the metallic region) drops anomalously as $T$ decreases.

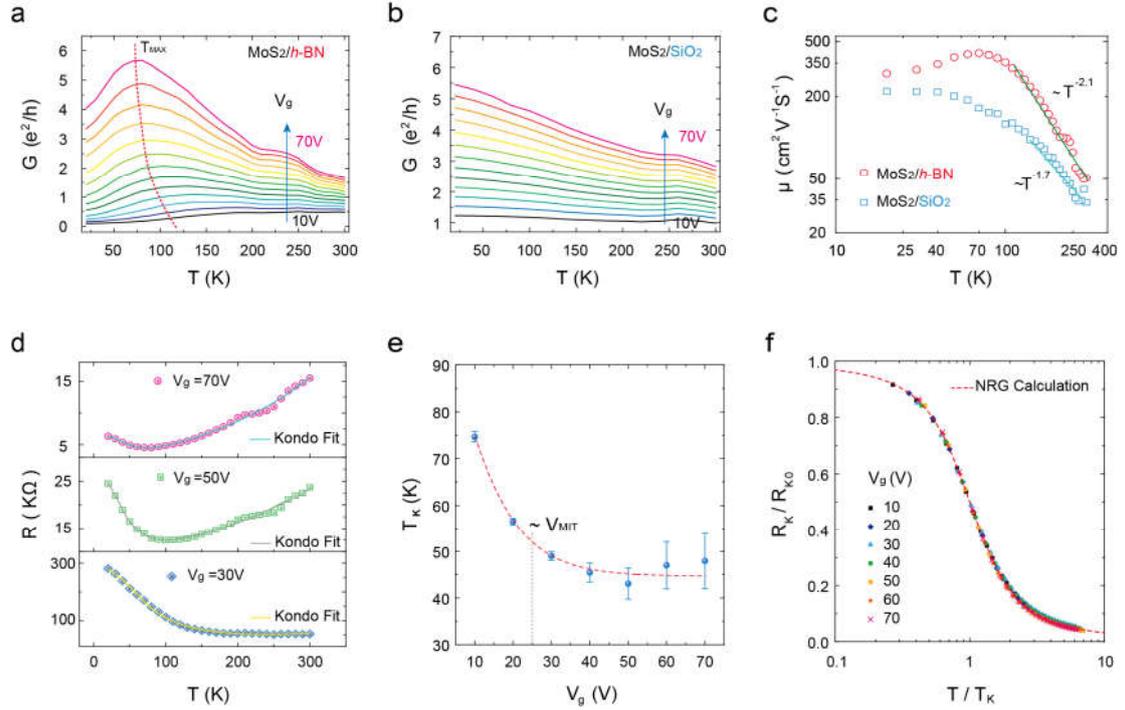

**Figure 3 | Carrier-density and temperature dependent properties of defective MoS$_2$ on *h*-BN substrate. a**, 4-probe electrical conductivity of MoS$_2$/*h*-BN devices as a function of $T$ and $V_g$. MIT can be observed when $G_{MIT} \sim e^2/h$. The anomalous peaks $T_{max}$ at low temperatures are marked out by the red dashed line. **b**, 4-probe electrical conductivity of MoS$_2$/SiO$_2$ devices as a function of $T$ and $V_g$. **c**, Temperature-dependent field effect mobility. The solid line shows the phonon-limited power law $\mu_{ph} \sim T^{-\gamma}$. $\gamma$ = 1.7 and $\mu \sim 200$ cm$^2$/V·s for MOS$_2$/SiO$_2$ sample. For MoS$_2$ on *h*-BN, $\gamma$ = 2.1 and $\mu$ reaches 405 cm$^2$/V·s. For $T <$ 100K, $d\mu_{FE}/dT > 0$ is observed for MoS$_2$/*h*-BN sample indicating an anomalous scattering mechanism. **d**, Temperature dependence of 4-probe resistance at $V_g$=70V, 50V and 30V for MoS$_2$/*h*-BN device, with the resistance minimum at 70K, 89K and 135K, respectively. **e**, Gate tuning of the Kondo temperature $T_k$. **f**, Normalized Kondo resistance versus normalized temperature for the data from $V_g$=10V to $V_g$=70V. The red solid line describes the universal Kondo behavior from numerical renormalization group calculations[31].

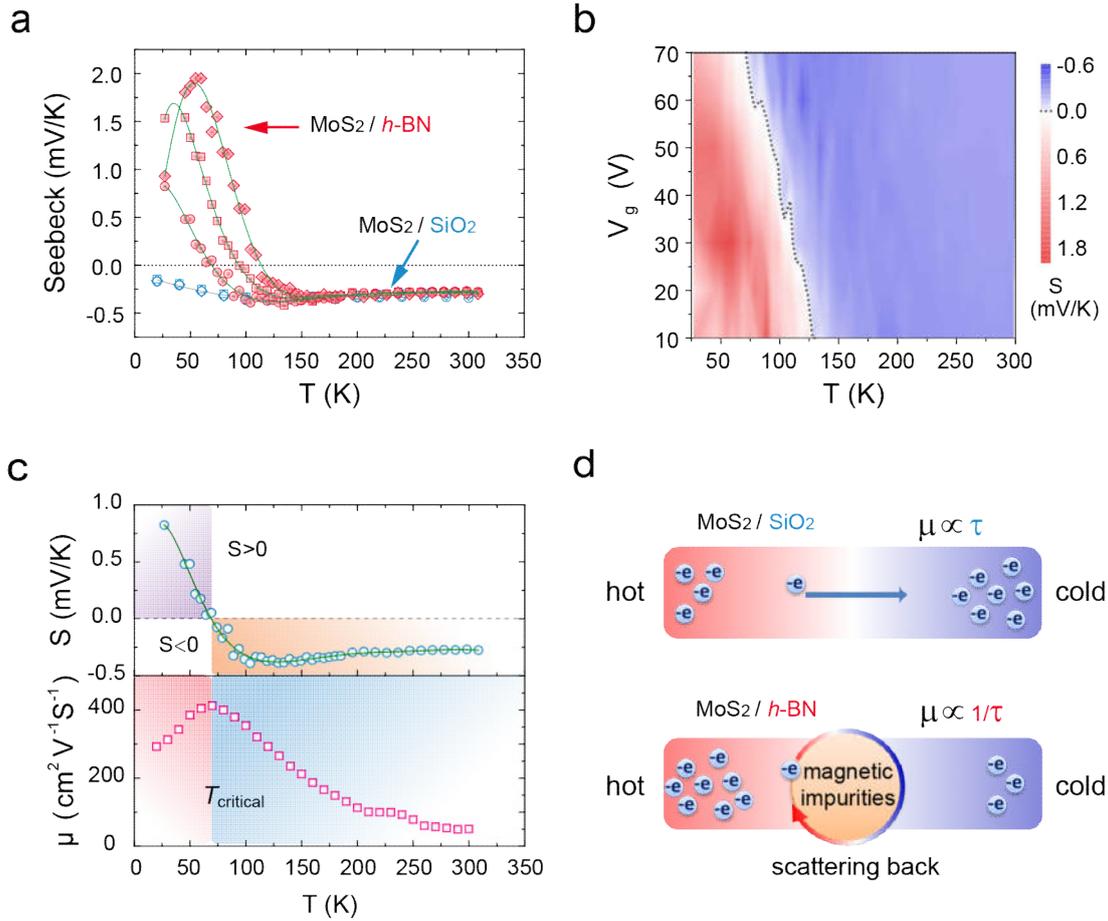

**Figure 4 | Thermoelectric transport and Seebeck coefficient measurement of defective MoS$_2$ on h-BN substrate. a**, Comparison of *S vs T* of MoS$_2$/SiO$_2$ and MoS$_2$/h-BN device at $V_g$=70V, 50V and 30V(from left to right). The solid lines guides the eye. MoS$_2$/SiO$_2$ sample shows negative values due to the diffusive type *n*-type charge carriers. For MoS$_2$/h-BN sample, at low temperatures, an anomalous sign-change in *S* occurs where the majority carriers are electrons. **b**, Colour contour plot of *S* values versus $V_g$ & *T* for MoS$_2$/h-BN device. The black dotted lineindicates the point with *S* ~ 0 mV/K. **c**, Seebeck coefficient and mobility as function of temperature at $V_g$=70V for MoS$_2$/h-BN device. **d**, Sketch for the electron diffusion in thermoelectric transport for MoS$_2$/SiO$_2$ sample and Kondo-driven MoS$_2$/h-BN sample. In MoS$_2$/SiO$_2$ sample, the conduction electrons will diffuse from the hot side to the cold side and the electron concentration is also modified by the variation in chemical potential due to the temperature gradient. For MoS$_2$/h-BN sample, due to the strong Kondo resonance, more conduction electrons are scattered back and accumulate at the hot side, thus producing a positive *S* in *n*-type MoS$_2$.

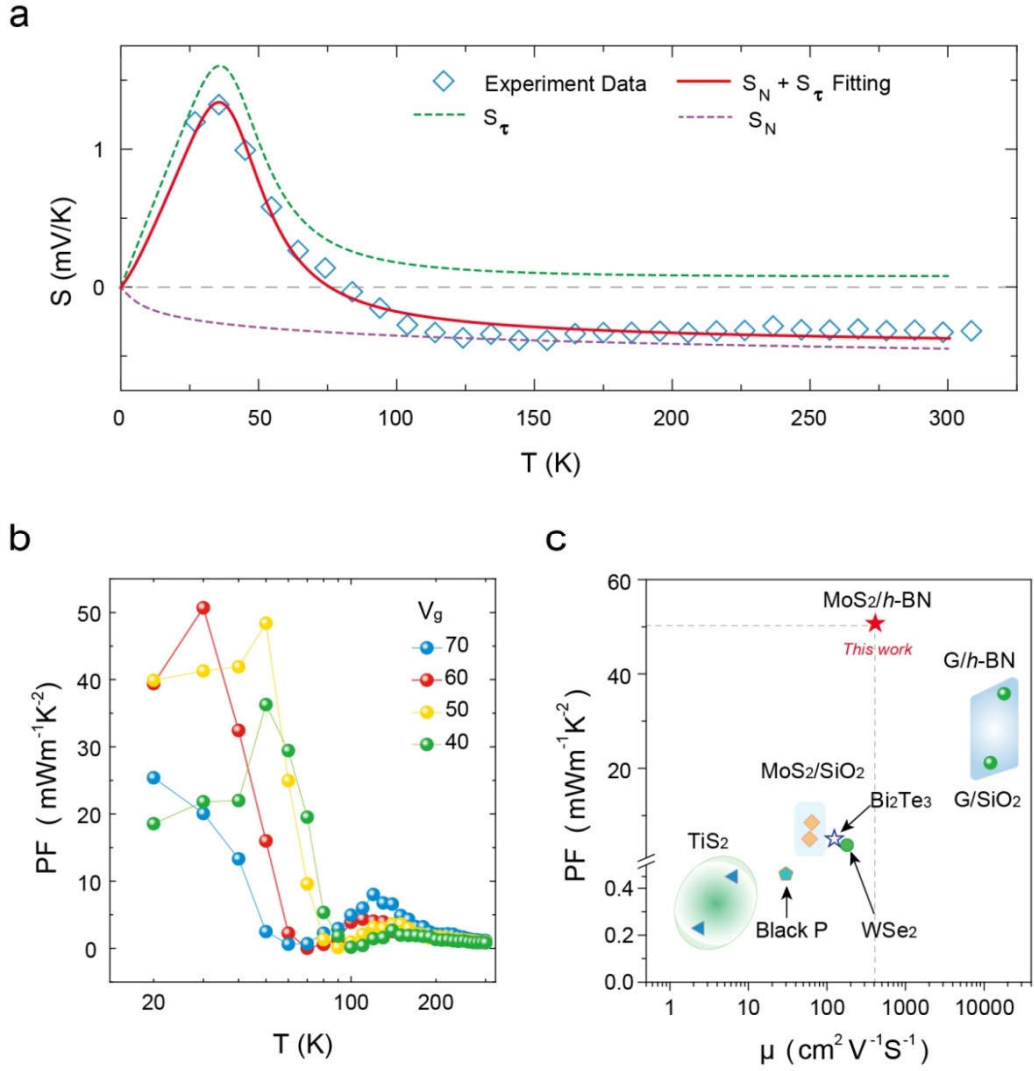

**Figure 5 | Thermoelectric performance of MoS$_2$/h-BN heterostructure. a**, Total $S$ values at $n= 2\times10^{12}$ cm$^{-2}$ are contributions from the energy dependent diffusive part, $S_N$ and the Kondo scattering part, $S_\tau$. At a fixed $n$, the total $S$ first exhibits a diffusive negative value at high temperatures from conducting electrons described by $S_N$. The $S_\tau$ part is negligible because the local magnetic moment is free while the system maximizes the entropy and weakens the scattering of the conduction electrons[48]. As the temperature decreases, the conventional diffusive contribution is weakened and the Kondo scattering term $S_\tau$ starts to dominate and shows large positive values. As temperature is decreased further, the impurities condensed into a singlet state and all the physical interactions start to freeze and the total $S$ goes back to zero as expected. **b**, Power factor as a function of temperature at different gate voltages. Additional Kondo induced peaks as high as 50 mW/mK$^2$ can be observed. **c**, The PF value of our Kondo-driven MoS$_2$/h-BN heterostructure shows superior thermoelectric performance compared to commercial thermoelectric material Bi$_2$Te$_3$ and other reported 2D materials. [G/SiO$_2$ & h-BN[41], MoS$_2$[32,33], WSe$_2$[42], TiS$_2$[43,44], Bi$_2$Te$_3$[45], BP[46]]

**Acknowledgements**

This research was supported by A*STAR Pharos Funding from the Science and Engineering Research Council (Grant No. 152 70 00015). C.W.Q acknowledges the A*STAR SERC Pharos grant no. 152 70 00014 with project no. R-263-000-B91-305. A.H.C.N acknowledges the NRF-CRP award R-144-000-295-281. We also thank Professor Zheng Y. for the helpful discussion and comments.

## Author Contributions

J.W., Y.P.L., K.P.L. and K.H. designed the experiments. J.W. and Y.L. performed the device fabrication and carried out the electric and thermoelectric measurement. Y.P.L. and J.W. performed the STM measurements. Y.Q. Cai and G.Z. carried out the DFT calculation. H.K.N. calculated the diffusive Seebeck coefficient. J.W., Y.P.L., K.H., K.P.L., Y.L., C.W.Q., J.T.L.T. and A.H.C.N. carried out data analysis and co-wrote the manuscript. K.W. and T.T. provided h-BN material. All authors commented on the manuscript.

## Additional Information

Correspondence and requests for materials should be addressed to K.H., K.P.L. and J.T.L.T.

## Competing financial interests

The authors declare no competing financial interests.